\documentclass[aps,pra,floatfix,twocolumn,nofootinbib,showpacs]{revtex4}
\usepackage{graphicx} \usepackage{amsmath} \usepackage{amssymb}
\newcommand{\BEQ}{\begin{equation}}
\newcommand{\EEQ}{\end{equation}}
\newcommand{\BEA}{\begin{eqnarray}}
\newcommand{\EEA}{\end{eqnarray}}

\renewcommand{\d}{{\rm d}}

\newcommand{\eps}{\varepsilon}

\newcommand{\A}{{\rm A}}
\newcommand{\B}{{\rm B}}
\newcommand{\C}{{\rm C}}
\renewcommand{\c}{C}

\renewcommand{\H}{{\cal H}}

\newcommand{\lb}{\langle\,}
\newcommand{\rb}{\,\rangle}
\newcommand{\tila}{\widetilde{\lambda}}
\newcommand{\tir}{\widetilde{r}}

\newcommand{\la}{\lambda}
\newcommand{\om}{\omega}

\newcommand{\lapr}{\lambda_{pr}}

\newcommand{\vb}{\vert\,}
\newcommand{\comment}[1]{}

\begin{document}

\title{Transferring elements of a density matrix}

\author{Armen E. Allahverdyan$^{1)}$ and Karen V. Hovhannisyan$^{1,2)}$ }

\address{
$^{1)}$
Yerevan Physics Institute, Alikhanian Brothers Street 2, Yerevan 375036, Armenia \\
$^{2)}$
Yerevan State University, A. Manoogian Street 1, Yerevan, Armenia}

\begin{abstract} We study restrictions imposed by quantum mechanics on
the process of matrix elements transfer. This problem is at the core of
quantum measurements and state transfer.  Given two systems $\A$ and
$\B$ with initial density matrices $\lambda$ and $r$, respectively, we
consider interactions that lead to transferring certain matrix elements
of unknown $\lambda$ into those of the final state ${\widetilde r}$ of
$\B$.  We find that this process eliminates the memory on the
transferred (or certain other) matrix elements from the final state of
$\A$.  If one diagonal matrix element is transferred, ${\widetilde
r}_{aa}=\lambda_{aa}$, the memory on each non-diagonal element
$\lambda_{a\not=b}$ is completely eliminated from the final density
operator of $\A$. Consider the following three quantities $\Re
\la_{a\not =b}$, $\Im \la_{a\not =b}$ and $\la_{aa}-\la_{bb}$ (the real
and imaginary part of a non-diagonal element and the corresponding
difference between diagonal elements). Transferring one of them, e.g.,
$\Re\tir_{a\not = b}=\Re\la_{a\not = b}$, erases the memory on two
others from the final state of $\A$. Generalization of these set-ups to
a finite-accuracy transfer brings in a trade-off between the accuracy
and the amount of preserved memory. This trade-off is expressed via
system-independent uncertainty relations which account for local aspects of
the accuracy-disturbance trade-off in quantum measurements.
\end{abstract}


\maketitle


\section{Introduction}
\label{intro}

Quantum mechanics imposes constraints on information processing.  Among
known examples of such contraints is the the fact that measuring an
unknown quantum state inevitably disturbs it. This fundamental feature
was known since the early days of quantum mechanics \cite{heisenberg},
and has been recently formalized via uncertainty relations and
information-disturbance trade-offs \cite{hoffmann,ozawa_p,ozawa,busch,lorenzo,martens};
see \cite{konrad,horod} for reviews.

Another constraint is the no-cloning theorem, which states that due to
linearity and unitarity of quantum dynamics there exists no physical
process that can produce perfect copies of a system that is initially in
an unknown quantum state \cite{wz}. The theorem is closely related to
the quantum measurement induced state-disturbance \cite{disturb}. There
are several important generalizations of the no-cloning theorem
\cite{yu,mor,jo,brd,superbrd,koashi_imoto,bh,dg,nodel,deuar,cloning_observable}.
\comment{, e.g.,
no-broadcasting principle \cite{brd,superbrd,koashi_imoto}, approximate
universal cloning \cite{bh}, probabilistic cloning \cite{dg}, and
no-deleting principle \cite{nodel}. }

Here we study limitations imposed by quantum mechanics on the process
of matrix elements transfer from one system to another. This problem
includes as particular cases quantum measurement and cloning (see below
for details). Before we formally pose the problem in the next section,
let us see where such transfer processes are encountered.

\subsection{Quantum measurement}

Let a quantum system is prepared in a (generally, unknown) state
described by a density matrix $\rho_{\rm S}$. For the measurement of an
observable $\hat{A}$ pertaining to the system quantum theory predicts that the
probabilities of observing various eigenvalues of $\hat{A}$ are given by
the Born rule:
\BEA
{\rm Pr}(\hat{A}=a)\,\equiv\,{\rm tr}[\rho_{\rm S}\hat{\Pi}_{\hat{A}}(a)],
\label{lend}
\EEA
where $\hat{\Pi}_{\hat{A}}(a)$ is the projector referring to the
eigenvalue $a$ of $\hat{A}$.

For describing the measurement process one has to include explicitly the
measuring apparatus, which|prior to its interaction with the system|is
in a {\it known} state with a density matrix $\rho_{\rm M}$. Several
requirements on $\rho_{\rm M}$ and the system-apparatus interaction are
to be satisfied by an {\it ideal} quantum measurement \cite{abn_pra}.
The basic|and in a sense minimal|requirement is that the initial
probabilities ${\rm Pr}(\hat{A}=a)$ in (\ref{lend}) are mapped to the
final probabilities of the apparatus observable $\hat{B}$
\cite{ozawa,ozawa_p,meas,neumann,abn_pra}:
\BEA
{\rm Pr}(\hat{A}=a)
 = {\rm Pr}_{\rm fin}(\hat{B}=a)\equiv {\rm tr}[\rho^{\rm fin}_{\rm M}\hat{\Pi}_{\hat{B}}(a)],
\label{bud}
\EEA
where $\rho^{\rm fin}_{\rm M}$ is the final (after interacting with the
system) density matrix of the apparatus, $\hat{\Pi}_{\hat{B}}(a)$ is the
projector of $\hat{B}$, and where for simplicity we assumed that
$\hat{A}$ and $\hat{B}$ have the same discrete spectra.

Eq.~(\ref{bud}) implies that the probabilities (\ref{lend}) of $\hat{A}$
can be obtained by looking at the statistics of the apparatus
observable $\hat{B}$. Relation (\ref{bud}) is satisfied with many
models of ideal quantum measurements
\cite{neumann,measurement_models,abn_pra}. It is supposed to hold for an
arbitrary initial density matrix $\rho_{\rm S}$, because the latter is
unknown.

Thus the quantum measurement means, in particular, transferring the
initial matrix elements of the tested system in the representation where
$\hat{A}$ is diagonal. The full transfer amounts to requiring
(\ref{bud}) for all independent probabilities. However, for concrete
purposes we can be interested only by certain probabilities ${\rm
Pr}_{\rm in}(\hat{A}=a)$ and then (\ref{bud}) is to be imposed only for
those probabilities.

For many models of quantum measurements it was observed that after
realizing an ideal measurements of the observable $\hat{A}$, the system
is left in a state with a density matrix diagonal in the
$\hat{A}$-representation \cite{neumann,measurement_models,abn_pra}. This
feature is closely related to the von Neumann projection
\cite{neumann,measurement_models,abn_pra}. It is now interesting to ask
what does happen to the state of the system after transferring ideally
the diagonal matrix elements according to (\ref{bud}), i.e., after satisfying
the minimal condition of quantum measurements.

\subsection{Polarization transfer}

Transfer of matrix elements is realized also in one of the
main methods of cooling, where polarization is transferred from one
system to another \cite{soren, exp_2,cooling}, e.g., from highly
polarized electron spins to almost unpolarized nuclear spins
\cite{cooling}. Polarization transfer is well known in NMR/ESR,
quantum/atomic optics, semiconductor physics, {\it etc} \cite{soren,
exp_2,cooling}. For the simplest example take two spin-$\frac{1}{2}$
density matrices for two systems
$$
\la=\frac{1}{2}[1+\vec{l}\,\vec{\sigma}], \qquad r=\frac{1}{2}[1+\vec{r}\,\vec{\sigma}],
$$
where $\vec{\sigma}$ are Pauli
matrices, and $\vec{l},\, \vec{r}$ are Bloch vectors.  Transferring
diagonal (non-diagonal) elements $\la_{11}=\tir_{11}$
($\la_{12}=\tir_{12}$) amounts to transferring the $z$ ($x$ and $y$)
component(s) of the Bloch vectors. Both these processes are well-studied
experimentally \cite{exp_1,soren,exp_2,cooling}.  Related processes of
energy (excitation) transfer are important in biological systems (e.g.,
photosynthesis) \cite{scholes}.  The energy transfer between two quantum
system means transferring the diagonal elements in the energy
representation.

\subsection{State transfer}

Quantum communication via (unknown) state transfer plays an
important role both for practical implementation of scalable quantum
processors and for understanding the efficiency of quantum computation;
see \cite{state_transfer_burg,state_transfer_chin} for reviews.  In many
theoretical studies devoted to the state-transfer problem
one simply assumes that the state of a finite-dimensional quantum
system (qubits or qutrits) is transferred to another system. Qubits and
qutrits can be understood literally as real systems with a finite number
of energy levels. However, more often than not, finite-dimensional
system are implemented in subspaces of a larger dimensional quantum
system; see \cite{knill} for a review. For instance, qubits can be
implemented via bosonic modes, or alternatively, they can be placed in
subspaces of a multi-qubit system, the purpose being immunization of the
qubit from decoherence (decoherence-free subspaces) of feasibility of
error-correcting schemes \cite{knill}.

Without going into details of implementation of qubits and qutrits in a
larger dimensional systems (see \cite{knill} in this context) one can
state that in all those cases where qubits and qutrits are not
understood literally, the resulting quantum state can be described via
suitable matrix elements of the full density matrix of the larger
(embodying) system; see \cite{exp_1} for experimental realizations.
Thus, in all those cases transferring quantum state
refers to certain (not all) elements of the full density matrix.

For various schemes of quantum state transfer it is of a clear
interest to understand what happens to the state of the source system
after the transfer has been realized, e.g., to what extent this final
state can serve as a source for another state transfer?

These examples show that transferring (certain) elements of the
(unknown) density matrix and understanding limitations imposed by
quantum theory on such processes is a relevant task.

The paper is
organized as follows. We formally state the studied problem in section
\ref{statement}.  The next two sections discuss limitation related to
the ideal transfer of matrix elements. In particular, section
\ref{diag-diag} discusses how the obtained results related to quantum
measurements.  Section \ref{nono-ido} describes a set-up for non-ideal
transfer processes. Details of such processes are presented in sections
\ref{nono-diag-diag} and \ref{nono-nondiag-nondiag}. We summarize in the
last section.

\section{Statement of the Problem}
\label{statement}

Consider a finite-dimensional quantum system $\A$. The information is
encoded into matrix elements of its density operator $\la$;
this situation is realized in the above examples. To be a
carrier of information this state has to be unknown. For simplicity
we assume that the state is {\it completely} unknown.

There is another,
composite system $\B+\C$ in some {\it known} state with density operator
$\om$. The Hilbert spaces of $\A$ and $\B$ have the same dimension: ${\rm
dim}\H_\A={\rm dim}\H_\B=N$. The initial state of the overall system
$\A+\B+\C$ is $\la\otimes\om$. Let $p,r=1,\ldots,N$ and
\BEA
\{|p\rb\}_{p=1}^n,\,\, \lb p\,|\,r\rb=\delta_{pr},\,\,\,\,\,\,
\{|\bar{p}\rb\}_{p=1}^n,\,\, \lb
\bar{p}\,|\,\bar{r}\rb=\delta_{pr},
\nonumber
\EEA
be two orthonormal bases in $\H_\A$ and $\H_\B$, respectively.
The interaction between $\A$ and $\B+\C$ is described by unitary
operator $U$. It will be chosen such that
for {\it any} initial density operator $\la$ of $\A$, certain
initial matrix elements $\la_{ab}=\langle a|\la|b\rangle$ of
\BEA
\la = {\sum}_{pr}\la_{pr}|p\rangle\langle r|
\label{katu}
\EEA
are equal to the corresponding matrix elements of the final state $\tir$ of $\B$:
\BEA
\la_{ab}=
\tir_{ab}=\langle \bar{a}|\tir|\bar{b}\rangle, \qquad \tir={\rm
tr}_{\A+\C} (U\,\la\otimes\om\, U^\dagger).\nonumber
\EEA
Here $\C$ is an auxiliary system (ancilla or environment).
After tracing it out, the considered dynamic operation
amounts to a trace-preserving
completely positive map acting on $\A+\B$.

We aim to understand implications of the matrix elements transfer from
$\A$ to $\B$ on the memory of the transferred elements $\la_{ab}$ (or
some other elements of $\la$) in the final state $\tila={\rm tr}_{\B+\C}
(U\,\la\otimes\om\, U^\dagger)$ of $\A$ (the formal definition of memory
is given in section \ref{nono-ido}).

Note that when all density matrix elements are transferred, the final
state of A cannot be equal to its initial state.  This follows from the
no-cloning theorem: there exists no quantum process that can produce
perfect copies of a system that is initially in an unknown quantum state
\cite{wz}. The theorem is closely tied to the fact that measuring the
unknown quantum state inevitably disturbs it \cite{disturb}.  However,
the no-cloning principle|even in the form of its various
generalizations
\cite{yu,mor,jo,brd,superbrd,koashi_imoto,bh,dg,nodel}| cannot be
applied directly to our problem, since here only certain (not all)
matrix elements are copied (transferred).

We choose the initial state of $\B+\C$ as
\BEA
\om=|\,\bar{1}\rb \lb \bar{1}\,|   \otimes|\,\c \rb\,\lb \c \,|,
\label{ini}
\EEA
where $|\c \rangle$ lives in the Hilbert space $\H_\C$ of $\C$. This
choice does not restrict generality provided that there are no
restrictions on the dimensionality of the Hilbert space ${\cal H}_{\rm
C}$ of ${\rm C}$, and provided that we are free to design unitary
evolutions for $\B+\C$. Indeed, an initial mixed state of $\B+\C$ can be
purified by extending $\C$ to a larger Hilbert space, while the
resulting pure state can be rotated to $|\,\bar{1}\rb\otimes|\,\c\rb$ by
a suitable unitary operator.

We represent the unitary operator $U$ as ($p=1,\ldots,N$)
\BEA
\label{riogo}
\label{unitar} U\,|p\rb\otimes |\,\bar{1}\rb\otimes|\,\c \rb ={\sum}_{k,l}
|k\rb\otimes|\,\bar{l}\rb\otimes|\,\c^{p}_{kl}\rb\equiv |\psi_p\rangle,
\EEA
where all summation indices run from $1$ to $N$, and where the vectors
$|\,\c^{p}_{kl}\rb$ with $p,k,l=1,\ldots,N$ live in $\H_\C$.

The unitarity of $U$ amounts to ($p,r=1,\ldots,N$)
\BEA
\label{bao}
\langle\psi_p|\psi_r\rangle=\delta_{rp}\qquad {\rm or}\qquad
{\sum}_{kl}\lb \c^p_{kl}\,|\,\c^{r}_{kl}\rb=\delta_{rp}.
\EEA

The final states
\BEA
\tila \quad {\rm and} \quad \tir=\sum_{a,b} \tir_{ab}|\bar{a}\rb\lb\bar{b}| \nonumber
\EEA
of $\A$ and $\B$, respectively, read from (\ref{unitar})
\BEA
\label{togo}
&&\tila={\sum}_{pr} \lapr \Theta_{pr},  \\
&&\tir_{ab} =
{\sum}_{pr}\lapr{\sum}_{k} \lb \c^r_{kb}|\,\c^p_{ka}\rb,
\label{kh2}
\EEA
where
\BEA
\label{kh1}
\Theta_{pr}\equiv {\sum}_{kn} |k\rb\,\lb n|{\sum}_{l}\lb
\c^r_{nl}|\,\c^p_{kl}\rb .
\EEA

The process of matrix elements transfer depends crucially on which
(diagonal or non-diagonal) elements are transferred. We therefore study
these cases separately.  Note that a diagonal density matrix $\la$ (with
unknown diagonal matrix elements) carries only a classical information.
Non-diagonal elements represent quantum aspects of the information
contained in the unknown state $\la$.

\section{ Diagonal to diagonal transfer: the ideal situation}
\label{diag-diag}

Assume that for every initial state $\la$ of $\A$ a diagonal element $\lambda_{aa}$ of $\A$ is transferred
to the diagonal element $\tir_{aa}$ of B:
\BEA
\label{vax}
\lambda_{aa}= \tir_{aa}.
\EEA
For this it is necessary to have [see (\ref{kh2})]
\BEA
\label{ba}
{\sum}_k\lb \c^r_{ka}|\,\c^p_{ka}\rb=\delta_{pr}\delta_{pa}
~~{\rm for ~ all~pairs}~(r,p).
\EEA
Eq.~(\ref{ba}) for $r=p=a$ implies
${\sum}_k\lb \c^a_{ka}|\,\c^a_{ka}\rb=1$. Combining this with
(\ref{bao}) under the same condition $p=r=a$ gives
$|\c^a_{kl}\rb=0$ for $l\not=a$. Eq.~(\ref{ba}) for $r=p=c\not=a$ gives
$|\c^c_{ka}\rb=0$ for every $c\not=a$. Altogether, we get
\begin{gather}
{\sum}_l\lb \c^a_{nl}\,\vb
\c^c_{kl}\rb=0~~{\rm for~every}~c\not=a~~{\rm or}\nonumber\\
\Theta_{a\not =c}=0,
\label{tu3}
\end{gather}
implying from (\ref{togo}, \ref{kh1}) that {\it due to transferring $\lambda_{aa}=
\tir_{aa}$ the memory on each initial non-diagonal element
$\lambda_{a\not =c}$ in the final density operator $\tila$ of $\A$ is
lost}; see (\ref{togo}).

Let us stress that the final state $\tila$ of $\A$ need not be diagonal
and that the memory on $\la_{aa}$ itself is conserved in $\tila$.  Note
that to be able to speak on the memory and its loss, we have to have
initially some freedom in choosing $\lambda_{a\not =c}$, i.e., the
latter should carry some information.

Recall from our introductory discussion that transferring the diagonal
elements is an essential part of the quantum measurement. The above
result on the memory loss of non-diagonal elements shows in which
specific sense the state of the measured system is disturbed after the
measurement. Studying disturbances induced by various quantum
measurement|in particular, studying the inevitable disturbance as a
function of the measurement accuracy|is a known subject; see
\cite{konrad,busch,lorenzo,martens,horod} for reviews. In particular, the
analysis of various models for the quantum measurement led to a
conclusion that after the ideal measurement is completed, the
post-measurement state is diagonal (an effect sometimes attributed to
decoherence) \cite{abn_pra,measurement_models}. It is seen from
(\ref{togo}, \ref{kh1}) and from (\ref{vax}--\ref{tu3}) that after the
ideal transfer of all diagonal matrix elements the final state of A need
not be diagonal, though it looses the memory on all non-diagonal
elements of the initial state of A. Indeed, assuming that all diagonal
are transferred we get from (\ref{vax}, \ref{tu3}) for the final state
$\tila$ of A:
\BEA
\langle s|\tila|t\rangle = {\sum}_p \lambda_{pp} \langle C^{p}_{tp}|C^{p}_{sp}\rangle.
\nonumber
\EEA
This means that the diagonalization of the post-measurement state was a
consequence of various additional conditions imposed on the quantum
measurement process; see \cite{abn_pra} for a detailed discussion.

To repeat, the basic (and minimal) requirement for the quantum
measurement is the transfer of diagonal matrix elements, and this
requirement leads to elemination of memory rather than to
diagonalization.

\section{ Transfer of non-diagonal elements.}
\label{nondiag-nondiag}

Demanding
\BEA
\label{karait1}
{\sum}_k \lb \c^r_{kb}|\,\c^p_{ka}\rb=\delta_{rb}\delta_{pa}~~{\rm for ~ all}~(r,p)
~{\rm and}~a\not=b,
\EEA
amounts to transferring ideally the corresponding non-diagonal element:
$$\tir_{ab}=\lambda_{ab}$$ for arbitrary initial state $\la$ of $\A$; see (\ref{togo}, \ref{kh1}).
The non-negativity of
${\sum}_k[\alpha^*\lb \c^a_{ka}|+\beta^*\lb \c^b_{kb}| \,]\,[\,\alpha|\,\c^a_{ka}\rb+\beta|\,\c^b_{kb}\rb]$
as a function of two complex numbers $\alpha$ and $\beta$ (Cauchy-Schwartz inequality) leads to
\begin{gather}
1={\sum}_k\lb \c^a_{ka}|\,\c^b_{kb}\rb\leq \sqrt{{\sum}_k \lb \c^a_{ka}|\,\c^a_{ka}\rb
{\sum}_k \lb \c^b_{kb}|\,\c^b_{kb}\rb },\nonumber\\
\label{jan}
\end{gather}
where the equality in (\ref{jan}) is due to (\ref{karait1}) under $r=a$ and $k=b$. The
inequality in (\ref{jan}) has to be saturated, since (\ref{bao}) implies
${\sum}_k\lb \c^a_{ka}|\,\c^a_{ka}\rb\leq 1$, ${\sum}_k\lb \c^b_{kb}|\,\c^b_{kb}\rb\leq 1$.
Thus we have ${\sum}_k\lb \c^a_{ka}|\,\c^a_{ka}\rb={\sum}_k\lb \c^b_{kb}|\,\c^b_{kb}\rb= 1$,
which together with (\ref{bao}) gives for any $k$
\BEA
\label{dedo}
|\c^a_{kl}\rb=0~~ {\rm for}~ l\not=a~~ {\rm and}~~
|\c^b_{kl}\rb=0~~ {\rm for}~ l\not=b.
\EEA
Eqs.~(\ref{togo}, \ref{kh1}, \ref{dedo}) lead to $\Theta_{a\not=b}=\Theta_{b\not=a}=0$, i.e.,
{\it the memory on the transferred non-diagonal element
$\lambda_{ab}$ in the final density operator $\tila$ is lost}; see (\ref{togo}).

Another consequence of saturating the inequality in (\ref{jan}) is that
$|\c^b_{kb}\rb=|\c^a_{ka}\rb$ for any $k$, which leads to
\BEA
{\sum}_l\langle \c^a_{nl}|\c^a_{kl}\rangle
=\langle \c^a_{na}|\c^a_{ka}\rangle
= {\sum}_l\langle \c^b_{nl}|\c^b_{kl}\rangle
=\langle \c^b_{nb}|\c^b_{kb}\rangle,\nonumber
\EEA
i.e., $\Theta_{aa}=\Theta_{bb}$, meaning that {\it memory on the difference of
diagonal elements $\lambda_{aa}-\lambda_{bb}$ in the final density
operator $\tila$ is lost}; see (\ref{togo}). Thus one ideal
nondiagonal-to-nondiagonal transfer eliminates the memory on three real
quantities, while one diagonal-to-diagonal ideal transfer eliminates
memory on $2(N-1)$ real quantities. The difference between these two
cases is that for the ideal nondiagonal-to-nondiagonal transfer the
memory on the transferred element itself is eliminated from the final state
of $\A$. This means that the non-diagonal elements (as compared to
diagonal ones) carry a different [more fragile] type of information.

Let us note that when only the real part of the non-diagonal element is
transferred, $\Re\,\tir_{ab}=\Re\,\lambda_{ab}$, for any initial density
matrix $\la$ of A, the above result on elimination of the memory on
$\lambda_{aa}-\lambda_{bb}$ still holds, while only the memory on the
imaginary part $\Im\lambda_{ab}$ is eliminated from the final density
operator $\tila$ of $\A$ (and {\it vice versa} when transferring the
imaginary part $\Im\,\tir_{ab}=\Im\,\lambda_{ab}$). Likewise,
transferring the difference between the eigenvalues,
$\tir_{aa}-\tir_{bb}=\la_{aa}-\la_{bb}$, eliminates the memory on
$\Im\lambda_{ab}$ and on $\Re\lambda_{ab}$. The derivation of these
facts is similar to that given around (\ref{karait1}--\ref{dedo}).
In this sense these three quantities $\la_{aa}-\la_{bb}$,
$\Im\lambda_{ab}$ and $\Re\lambda_{ab}$ are complementary to each other.

It is seen that transferring an eigenvalue $\tir_{aa}=\la_{aa}$ implies
different (more severe) consequences for the memory of non-diagonal
elements, than transferring an eigenvalaue difference
$\la_{aa}-\la_{bb}$. Nevertheless, when all $N-1$ independent diagonal
elements are transferred either directly, or via their differences, the
resulting damage to the memory of non-diagonal elements is the same,
i.e., the memory on all non-diagonal elements is erased. For the direct
transfer this is obvious from (\ref{tu3}), while for the second situation of
transferring the eigenvalue differences this follows from the fact that
$\tir_{aa}-\tir_{bb}=\la_{aa}-\la_{bb}$ implies conditions (\ref{dedo}).

\section{Non-ideal transfer and a measure of memory}
\label{nono-ido}

While the above results refer to the ideal transfer, it is important to
see how much memory can be preserved under a non-ideal, {\it
finite-accuracy} transfer. Naturally, the general purpose of studying
non-ideal transfer is to find some compromise between transferring
diagonal elemenens and erasing the memory of non-diagonal elements in
the final state of A.

First let us recall an obvious fact that when transferring (ideally or
not) diagonal elements (i.e., positive numbers summing to one), we have
to describe the transfer of independent diagonal elements only.

Now if the ideal transfer corresponds to $\tir_{aa}^{\rm [id]}=\lambda_{aa}$, its
non-accurate version is defined to be
\BEA
\label{bordeau}
\tir_{aa}=\varepsilon_a\,\lambda_{aa},
\EEA
where we assume that $\varepsilon_a$ does not depend on the initial
state $\la$, and where $\frac{\tir_{aa}^{\rm
[id]}-\tir_{aa}}{\tir_{aa}^{\rm [id]}} =1-\varepsilon_a$ varies between
zero and one, $0<1-\varepsilon<1$, and characterizes the relative
accuracy of the transfer. (Clearly, one cannot have $\varepsilon_a >1$,
because the positive diagonal elements should sum to one for all initial
state $\la$; we also recall that (\ref{bordeau}) is demanded for
independent probabilities only.)
The notion of the relative accuracy is frequently met in
the standard analysis of experimental errors \cite{taylor}.

If $\lambda_{aa}$ is considered as a signal, $\varepsilon_a<1$
corresponds to reducing (by a fixed amount) the signal magnitude without introducing any bias. If
some noise is present during the actual transfer of the matrix element,
this reduction will correspond to decreasing the signal-to-noise ratio,
because weaker signals are more difficult to detect \cite{taylor}.

Conditions (\ref{bordeau}) are to be imposed on independent
probabilities only, so that at best we can have only $N-1$ such
constraints.

Note that (\ref{bordeau}) is certainly not the only way of defining
non-ideal measurements. For instance, in the literature devoted to
quantum measurements one sometimes employs the Heisenberg representation
\cite{ozawa,arthurs}. Within this representation there is a reasonable
definition of non-ideality, which is related to considering Heisenberg
operators as signals \cite{ozawa,arthurs}. In particular, the Heisenberg
operator of the apparatus variable after the system-apparatus
interaction is compared to the system-variable Heisenberg operator
before this interaction \cite{ozawa}. Other approaches to non-ideal
measurements are reviewed in \cite{martens,konrad,horod}.

However, condition (\ref{bordeau}) seems to be the simplest possibility
(at least within the employed Schroedinger representation) for
introducing a finite non-accuracy without introducing any bias.

\subsection{Quantifying the memory}
\label{memo}

The memory on the initial non-diagonal element $\la_{a\not= c}$ in the
final state (\ref{togo}) is most naturally quantified by checking the
response of the final state to perturbations in $\la_{a\not= c}$.  We
take another initial state $\la'$ of A, such that all matrix elements of
$\la$ and $\la'$ are identical besides the real and/or imaginary part of
$\la_{a\not= c}$.  Naturally, such a $\la'$ can always be found, due to
the basic constraint on $\la_{a\not= c}$: $|\la_{a\not=
c}|\leq\la_{aa}\la_{cc}$. (If $\la_{aa}=0$ (or $\la_{cc}=0$), the very
freedom in choosing $\la_{a\not= c}$ is absent, so there is no point in
discussing its memory loss.)

Provided that the (small) difference between $\la$ and $\la'$ is fixed, we look
at the difference between the corresponding final states $\tila$ and
$\tila'$. This amounts to taking the derivatives
${\partial\tila}/{\partial \,\Re\la_{a
c}}|_{\Im\la_{ac}}$ and ${\partial\tila}/{\partial
\,\Im\la_{ac}}|_{\Re\la_{ac}}$, which quantify, respectively, the
memory on the real and imaginal parts of $\la_{ac}$. These are still
matrices, but the strength of the dependence of $\tila$ on
$\Re\la_{a\not=c}$ or on $\Im\la_{a\not=c}$
can be characterized via norms $||{\partial\tila}/{\partial \,\Re\la_{a
c}}|_{\Im\la_{ac}}\,||$ and $||{\partial\tila}/{\partial
\,\Im\la_{ac}}|_{\Re\la_{ac}}\, ||$. Since all norms are
equivalent in a finite-dimensional Hilbert space|i.e.,
given two norms $||.||_1$ and $||.||_2$, there exist positive constants
$a$ and $b$ such that $a||A||_2 \leq ||A||_1 \leq b ||A||_2$ for any
matrix $A$|we work with the Euclidean norm
\BEA
||A||\equiv \sqrt{{\rm tr}(AA^\dagger)},
\label{bala}
\EEA
where $A^\dagger$ is the hermitean conjugate of $A$.  Finally, the
memory of $\tila$ on $\la_{a\not=c}$ (i.e., on {\it both}
$\Re\la_{a\not=c}$ and $\Im\la_{a\not=c}$) is defined as
\BEA
\frac{1}{2}
\sqrt{||{\partial\tila}/{\partial \,\Re\la_{a c}}||^2 +
||{\partial\tila}/{\partial \,\Im\la_{a c}}||^2}=||\Theta_{a\not=c}||,
\label{navukhodonosor}
\EEA
where $\frac{1}{2}$ is introduced for convenience, and where $\Theta_{ac}$ is defined in (\ref{kh1}).

That the memory of $\tila$ on (the real and imaginary parts of) $\la_{ac}$ can be characterized by
$||\Theta_{ac}||$ is verified also by studying the matrix gradient of
$\tila$, whose modulus is limited by $||\Theta_{a\not=c}||$ and
$\frac{1}{\sqrt{2}}||\Theta_{a\not=c}||$ from above and below,
respectively \cite{gradient}.

Note that in the initial state $||\Theta_{ac}||= 1$ (perfect memory),
while after a trace-preserving completely positive map $\lambda\to
\tila$, we get that the memory on a matrix element can only decrease
$||\Theta_{ac}||\leq 1$. We skip the derivation of this fact, because it
is very similar to the derivation presented around
(\ref{grom1}--\ref{grom3}). Now assume that after transferring matrix
elements, when A has reached the state $\tila$, the system A is subjected to a
closed-system dynamics: $\tila\to\hat{U}\tila\hat{U}^\dagger$, where
$\hat{U}$ is a unitary operator living in the Hilbert space of A, and
generated by the free Hamiltonian of A. Physically, this means that
there is time-lag between realizing the matrix elements transfer and
checking for memory. Now as follows from the unitary invariance of the
norm (\ref{bala}), $||A||=||\hat{U}\,A\,\hat{U}^\dagger||$, the memory
on a non-diagonal matrix element will not change under a local
(closed-system) dynamics.

It is thus seen that the introduced measure of memory does have desired
features that support its interpretation. The above reasoning can be
applied to quantifying the memory on various combinations of matrix
elements; see below.

\subsubsection{Fidelity}
\label{m_vs_f}

Note that for describing the state disturbance during quantum
measurements and cloning one frequently employs the fidelity between the
final and initial state; see, e.g., \cite{konrad,horod,brd}. For our
situation this implies that for quantifying disturbances in the state of
A, we try to use the fidelity $F(\lambda,\tila)$ between the initial
$\lambda$ and final $\tila$ states of A:
\BEA
F(\lambda,\tila)=\left({\rm tr}\sqrt{\left [\,\lambda^{1/2}\tila\, \lambda^{1/2}     \,\right]}\right)^2.
\nonumber
\EEA
Features of the fidelity are reviewed in \cite{jojo}. In particular,
$F(\lambda,\tila)$ varies between $0$ and $1$ and it is equal to $1$ if
and only if $\lambda=\tila$. Thus, its deviation from $1$ is supposed to
quantify the "distance" between $\lambda$ and $\tila$. The largest
"distance" $F(\lambda,\tila)=0$ is achieved for orthogonal states
$\lambda$ and $\tila$.

We saw above that the memory on non-diagonal elements disappeared after
the diagonal elements transfer. This naturally means that the final state of A
differs from its initial state, and thus the fidelity is smaller than
one. The converse is clearly not correct: the fidelity strictly smaller
than one yet does not imply the specific memory loss effect found above.
In other words, for the present problem the global measures of the state
disturbance (such as the fidelity) are not adequate, because they can
hide important physics. We need a local description of the
disturbances induced in the final state of the source system A, such as
the measure of memory introduced above.

Looking at the situation from a different angle, let us note the
following undesirable feature of the fidelity (as would-be employed for
the present situation). At the end of section \ref{memo} we noted that
the introduced measure of memory is invariant with respect to unitary
(closed-system) dynamics. This is clearly not the case with the
fidelity, because in general
$F(\lambda,\tila)\not=F(\lambda,\hat{U}\,\tila\hat{U}^\dagger)$ for a unitary $\hat{U}$.
We note in this context that a clear
analysis of various general drawbacks of the fidelity is presented in
Ref.~\cite{deuar}.

\section{Diagonal to diagonal transfer: non-ideal situation.}
\label{nono-diag-diag}

We shall study
the maximal possible memory on the initial non-diagonal elements
$\la_{a\not=c}$ under a finite-accuracy transfer (\ref{bordeau}). It proves
more convenient to assume $N\geq 3$ and to start immediately with the
simultaneous non-ideal transfer of two (independent) diagonal elements of the $N\times
N$ density matrix:
\begin{gather}
\tir_{aa}=\varepsilon_a\,\lambda_{aa}, ~~ \tir_{bb}=\varepsilon_b\,\lambda_{bb}, ~~ 0<\varepsilon_a<1,
~~ 0<\varepsilon_b<1,\nonumber\\
\label{dd}
\end{gather}
where $\varepsilon_a$ and $\varepsilon_b$ do not depend on the initial
state $\la$ and quantify the non-ideality.  This case is generic, since
the non-ideal transfer of one (or several) elements can be recovered
from it; see below. (For $N=2$ we have only $\tir_{aa}=\varepsilon_a\,\lambda_{aa}$ instead of (\ref{dd}).)
Instead of (\ref{ba}) we get from (\ref{dd})
\begin{gather}
{\sum}_k\lb \c^r_{ku}|\,\c^p_{ku}\rb=\varepsilon_u\delta_{pr}\delta_{pu}
~{\rm for~ all}~ (r,p)~{\rm and}~u=a,b. \nonumber\\
\label{aba}
\end{gather}
Eq.~(\ref{aba}) for $r=p\not=a$ and for $r=p\not=b$ gives for any $k$
\BEA
\label{babo}
|\c^p_{ka}\rb=0~~ {\rm for}~ p\not=a~~ {\rm and}~~
|\c^p_{kb}\rb=0~~ {\rm for}~ p\not=b.
\EEA

Given (\ref{dd}, \ref{aba}, \ref{babo}) we now establish an upper bound on $||\Theta_{a\not=c}||$.  Let
us define
\BEA
z^{r~p}_{nl\,kl}\equiv \lb \c^r_{nl}|\,\c^p_{kl}\rb ,
\EEA
and let ${\sum}'_{l}$
be the summation over $l=1,\ldots,N$ excluding $l=a$ and $l=b$. We get
from (\ref{togo}, \ref{kh1})
\BEA
\label{grom1}
||\Theta_{a\not=c}||^2\equiv {\sum}_{k,n}\left|{\sum}_l z^{c~a}_{nl\,kl}\right|^2
\leq {\sum}_{k,n}\left[{\sum}_l |z^{c~\,a}_{nl\,kl}|\right]^2\\
\label{grom2}
\leq {\sum}_{k,n}\left[{\sum}'_l \sqrt{z^{c~\,c}_{nl\,nl}}\sqrt{z^{a~a}_{kl\,kl}}  \right]^2~~~\\
\leq {\sum}_n{\sum}'_l z^{c~\,c}_{nl\,nl}\,\,\, {\sum}_k{\sum}'_l z^{a~a}_{kl\,kl},~~~
\label{grom3}
\EEA
where the inequalities in (\ref{grom2}) and (\ref{grom3}) are due to the Cauchy-Schwartz
inequality, while in (\ref{grom2}) we additionally used (\ref{babo}). We
now get from (\ref{grom3}) and (\ref{bao}, \ref{aba}, \ref{babo})
\BEA
\label{odzin1}
&&||\Theta_{a\not=b}||\leq \sqrt{(1-\varepsilon_a)(1-\varepsilon_b)},~~~~\\
\label{odzin2}
&&||\Theta_{a\not=c}||\leq \sqrt{(1-\varepsilon_a)}~~{\rm for~every}~~ c\not =a,\, c\not =b.~~~~\\
&&||\Theta_{b\not=c}||\leq \sqrt{(1-\varepsilon_b)}~~{\rm for~every}~~ c\not =a,\, c\not =b.~~~~
\label{odzin3}
\EEA
These inequalities|which are akin to the uncertainty relations|relate the
non-ideality of transfer to the maximal possible amount of the
conserved memory. Note that the bound on $||\Theta_{a\not=b}||$ is tighter than those on $||\Theta_{a\not=c}||$
and $||\Theta_{b\not=c}||$: once the diagonal elements $\lambda_{aa}$ and $\lambda_{bb}$ are transferred,
the memory o the cross-non-diagonal element $\lambda_{ab} $ is the most vulnerable one.

The extension of (\ref{odzin1}, \ref{odzin2}) to transferring
non-ideally several matrix elements should be obvious, since the
non-diagonal elements under such a transfer fall naturally into two
classes, which correspond to (\ref{odzin1}) and (\ref{odzin2},
\ref{odzin3}) respectively.

Let us show that the bounds (\ref{odzin1}, \ref{odzin2}) are
saturated by the proper choice of $|\c^p_{kn}\rb$. To this end assume
that ${\rm dim}\H_\C=1$: $|\,\c^p_{kb}\rb=\c^p_{kb}|\,\c\rb$, where $\c^p_{kb}$
are c-numbers satisfying (\ref{bao}). Thus we study a unitary interaction between A and B.
Choosing for $N=3$
\begin{gather}
\label{shalom1}
\c^1_{11} = \sqrt{\varepsilon_1}, ~~~~
\c^1_{13} = \sqrt{1-\varepsilon_1},~~~~
\c^2_{22} = \sqrt{\varepsilon_2}, \\
\c^2_{23} = \sqrt{1-\varepsilon_2}, ~~~~~ \c^3_{33} = 1
\label{shalom2}
\end{gather}
while all other $\c^p_{kb}$ with $p,k,b=1,2,3$ are zero, we satisfy the
unitarity conditions (\ref{bao}) and realize the optimal
memory-conserving non-ideal transfer (\ref{dd}) with $a=1$ and $b=2$.
Now (\ref{odzin1}, \ref{odzin2}) become equalities.

\section{ Non-ideal transfer of non-diagonal elements}
\label{nono-nondiag-nondiag}

Let us now turn to a finite-accuracy, non-diagonal-to-non-diagonal
transfer
\BEA
\label{miller}
\tir_{ab}=\eta\lambda_{ab}, \qquad
a\not =b \quad {\rm and}\quad  0<|\eta|<1,
\EEA
where $\eta$ can be a complex number, and where $|\eta|$ characterizes the accuracy
in the same sense as $\varepsilon_a$ in (\ref{bordeau}).
We shall find out how the
memory in the non-diagonal element $||\Theta_{ab}||$ and the memory
$||\Theta_{aa}-\Theta_{bb}||$ on the difference betwen the diagonal
elements are bounded. Initially, we restrict ourselves to finding the maximal
possible memories for the c-number case
\BEA
\label{taran}
|\c^p_{kb}\rb=\c^p_{kb}|\,\c\rb.
\EEA
Already this particular case will allow us to draw general conclusions
on the difference with the non-ideal diagonal-to-diagonal transfer. More
general cases will be discussed below.

For (\ref{miller}) to hold for an arbitrary initial state $\la$ of $\A$ we need
\BEA
\label{varaz}
{\sum}_k \c^r_{kb}\c^{p\,\,*}_{ka}=\eta\delta_{rb}\delta_{pa}~~{\rm for ~ all}~(r,p)
~{\rm and}~a\not=b.
\EEA
This implies ${\sum}_k \c^a_{kb}\c^{a\,\,*}_{ka}={\sum}_k \c^b_{kb}\c^{b\,\,*}_{ka}=0$,
and then
\begin{gather}
\label{kom1}
||\Theta_{a\not=b}||^2
=\phi^{a}_{a}\phi^{b}_{a}+\phi^{a}_{b}\phi^{b}_{b}+\Lambda_{ab},\\
\label{rostov}
\phi^{u}_{v}\equiv {\sum}_k |\c^u_{kv}|^2,  \\
\Lambda_{a\not=b}\equiv {\sum}'_{[sl]}\left[{\sum}_k \c^a_{kl}\c^{a\,\,*}_{ks}\right]
\left[{\sum}_n \c^b_{ns}\c^{b\,\,*}_{nl}\right],
\label{kom2}
\end{gather}
where ${\sum}'_{[sl]}$ means that the four pairs $(s,l)=(a,a), (a,b), (b,a), (b,b)$
are excluded from the summation over $s=1,\ldots,N$ and $l=1,\ldots,N$.
In estimating $|\Lambda_{a\not=b}|$ from above we proceed by applying the Cauchy-Schwartz inequality
and using (\ref{kom1}):
\BEA
|\Lambda_{a\not=b}|\leq
{\sum}'_{[sl]}\left[{\sum}_k |\c^a_{kl}||\c^{a\,\,*}_{ks}|\right]
\left[{\sum}_n |\c^b_{ns}||\c^{b\,\,*}_{nl}|\right]\\
\leq {\sum}'_{[sl]} \sqrt{ \phi^{a}_{l} \phi^{a}_{s} \phi^{b}_{l} \phi^{b}_{s}   }
\leq \sqrt{
{\sum}'_{[sl]}\phi^{a}_{l} \phi^{a}_{s} {\sum}'_{[sl]}\phi^{b}_{l} \phi^{b}_{s}
}.
\label{dodosh3}
\EEA
Working out (\ref{dodosh3}) and combining it with (\ref{kom1}) we obtain
\BEA
\label{bern1}
||\Theta_{a\not=b}||^2
&\leq& \phi^{a}_{a}\phi^{b}_{a}+\phi^{a}_{b}\phi^{b}_{b}~~~~~~
\\
&+&
\sqrt{[1-(\phi^a_a+\phi^a_b)^2]  [1-(\phi^b_a+\phi^b_b)^2]  }\equiv F,~~~~~
\label{bern2}
\EEA
where we used $\sum_k\phi^r_k=1$; see (\ref{rostov}, \ref{bao}).
We now maximize $F$ in the RHS of (\ref{bern2}) so as to obtain a bound
on $||\Theta_{a\not=b}||^2$ that holds for any $\{\c^b_{kl}\}$. The
maximization is carried out under two constraints: {\it i)}
$\phi_a^a\phi_b^b\geq |\eta|^2$, which follows from applying the
Cauchy-Schwartz inequality to (\ref{varaz}) with $r=b$ and $p=a$; {\it
ii)} $\phi_a^a+\phi_b^a\leq 1$ and $\phi_b^b+\phi_a^b\leq 1$, which
follow from the unitarity condition (\ref{bao}). Note from (\ref{bern1}, \ref{bern2})
that the maximum of $F$ over $\phi^b_a$ can be reached only at the
boundaries of its range, i.e., at $\phi^b_a=0$ or at
$\phi^b_a=1-\phi_b^b$. The same holds for $\phi^a_b$. Direct
inspection shows that the maximum of $F$ is reached for
$\phi^b_a=\phi^a_b=0$ and $\phi_a^a=\phi_b^b=|\eta|$:
\BEA
\label{opus}
||\Theta_{a\not=b}||\leq\sqrt{1-|\eta|^2}.
\EEA
Comparing (\ref{opus}) with (\ref{odzin2}) we see that the maximal
amount of the preserved memory on the non-diagonal element is larger for
the non-ideal nondiagonal-to-nondiagonal transfer than for the
diagonal-to-diagonal transfer with the same degree of non-ideality.

For the transfer $\tir_{21}=\eta\lambda_{21}$ and for $N=2$
the bound (\ref{opus}) is saturated by the following choice of
$\{\c^b_{kl}\}$
\BEA
\label{un}
&&\c^1_{11} = 1,~~~\c^1_{21}=\c^1_{12}=\c^1_{22} = 0,\\
&&\c^2_{21} = \sqrt{1-|\eta|^2},~~ \c^2_{12} = \eta, ~~\c^2_{22}=\c^2_{11}=0,
\label{bun}
\EEA
with an obvious generalization to $N\geq 3$. For the example (\ref{un}, \ref{bun})
let us write down the final states of $\A$ ($\tila$) and $\B$ ($\tir$):
\BEA
\label{sh}
\tila=
\left (\begin{array}{rr}
 \la_{11}+\la_{22}|\eta|^2  & ~~\la_{12}\sqrt{1-|\eta|^2}  \\
 \la_{21}\sqrt{1-|\eta|^2}  & ~~\la_{22}(1-|\eta|^2) \\
\end{array}\right ),\\
\tir=
\left (\begin{array}{rr}
 \la_{11}+\la_{22}(1-|\eta|^2)  & ~~\eta^*\la_{12}  \\
 \eta\la_{21}~~~~~~~~~~~~   & ~~\la_{22}|\eta|^2 \\
\end{array}\right ).
\EEA
Eq.~(\ref{sh}) shows that for a very inaccurate
non-diagonal-to-non-diagonal transfer $|\eta|\ll 1$, the disturbance
introduced in the final state of $\A$ can be a higher-order effect,
$\propto |\eta|^2$, i.e., in the {\it perturbative sense} the
disturbance can be neglected. This effect is clearly impossible for the
inaccurate diagonal-to-diagonal transfer. There for a small $\eps$ the
induced disturbance is at least of order $\eps$; see
(\ref{odzin1}--\ref{odzin3}). An explanation of this difference is that for the
diagonal-to-diagonal transfer the accuracy factor $\eps$ is strictly
non-negative. So after the zero-order term $1$ in the
memory-disturbance factor one can have a first-order term proportional
to $\eps$; see (\ref{odzin1}--\ref{odzin3}). In contrast, for the
non-digonal-to-non-diagonal transfer the accuracy factor $\eta$ is
generally complex; thus the first-order factor $|\eta|$ cannot
appear (since it is not smooth with respect to $\Re\eta$ and
$\Im\eta$), and the expansion starts from the second-term $|\eta|^2$.

It remains to see what happens to the memory on the diagonal element
difference $\la_{aa}-\la_{bb}$ under non-ideal transfer (\ref{miller}).
This memory is quantified by
\BEA
\frac{1}{\sqrt{2}}\Vert\Theta_{aa}-\Theta_{bb}\Vert,
\label{balasan}
\EEA
where the factor $\frac{1}{\sqrt{2}}$ is introduced for convenience. The
suitability of this memory measure can be argued for in the same way as
after (\ref{navukhodonosor}). In particular, (\ref{balasan}) is equal to
its maximal value $1$ in the initial state.

We now have:
\BEA
\Vert\Theta_{aa}-\Theta_{bb}\Vert^2 =
\Vert\Theta_{aa}\Vert^2+\Vert\Theta_{bb}\Vert^2 - 2{\rm
tr}\left(\Theta_{aa}\Theta_{bb}\right)
\label{zz}
\EEA
Recalling that we are restricted to the c-number situation $|\c^p_{kb}\rb=\c^p_{kb}|\,\c\rb$,
and denoting $\mu_{ls}^{pr}\equiv{\sum}_n{C^p_{nl}C^{r\, *}_{ns}}$ we get
\BEA
{\rm tr}\left(\Theta_{aa}\Theta_{bb}\right)={\sum}_{ls}{|\mu^{ab}_{ls}|^2}\geq |\mu^{ab}_{ab}|^2=|\eta|^2,
\label{glan}
\EEA
where we employed (\ref{varaz}) in the last equality. Combining (\ref{zz}, \ref{glan})
with $\Vert\Theta_{aa}\Vert\leq 1$ and $\Vert\Theta_{bb}\Vert\leq 1$, we get
\BEA
\label{gajl}
\label{gail1}
\frac{1}{\sqrt{2}}
\Vert\Theta_{aa}-\Theta_{bb}\Vert
\leq
\sqrt{1-\eps^2}.
\EEA
This is the sought upper bound on the memory of $\la_{aa}-\la_{bb}$.
It has the same form as (\ref{opus}).

For $N=3$ this limit is reached by a unitary in the first three rows of
which only the following elements are nonzero
\BEA
C^1_{11}=1,\;C^2_{12}=\eps,\;C^2_{32}=\sqrt{1-\eps^2},\;C^3_{13}=1
\EEA
This example shows a general property of the nullification of
$\Vert\Theta_{ab}\Vert$ in the regime where $\frac{1}{\sqrt{2}}\Vert\Theta_{aa}-\Theta_{bb}\Vert$ is in its maximum (we omit
the formal proof of this statement). However,
the maximization of $||\Theta_{ab}||$ does not nullify
$\frac{1}{\sqrt{2}}\Vert\Theta_{aa}-\Theta_{bb}\Vert$. The example (\ref{un}, \ref{bun}) illustrates this
fact since it leads to $\frac{1}{\sqrt{2}}\Vert\Theta_{aa}-\Theta_{bb}\Vert=1-\eps^2$, which is naturally smaller
than the optimal bound (\ref{gail1}).

\subsection{Extending the bound (\ref{opus}) to more general situations   }

In obtaining the bounds (\ref{opus}, \ref{gail1}) we constrainted
ourselves by (\ref{taran})|i.e., by one-dimensional Hilbert space ${\cal
H}_{\rm C}$, which amounts to a unitary interaction between A and
B|because so far we were not able to get more general analytic results.
It is interesting to know whether taking larger dimensions of ${\cal
H}_{\rm C}$ can improve the bounds (\ref{opus}, \ref{gail1}).  This
question was studied numerically for several values of ${\rm dim}{\cal
H}_{\rm C}$ and ${\rm dim}{\cal H}_{\rm A}={\cal H}_{\rm B}$. We imposed
condition (\ref{miller}) and numerically maximized the memories over the
available unitary transformations. The standard optimization routine
{\sl NMaximize} of Mathematica 7 has been employed.  Our numerical
results fully confirmed the bounds (\ref{opus}, \ref{gail1}); see Tables
I and II. We conjecture that these bounds hold for arbitrary values of
${\rm dim}\,{\cal H}_{\rm C}$.

\begin{table}
\caption{The maximal value of $||\Theta_{a\not=b}||^2$ for ${\rm dim}{\cal H}_{\rm A}={\cal H}_{\rm B}=3$ and various values
of ${\rm dim}{\cal H}_{\rm C}$ and the non-ideality parameter $\varepsilon$. The numerical results were obtained via running
the {\sl NMaximize} routine of Mathematica 7 for 37 iterations. The values for $||\Theta_{a\not=b}||^2$ are close to the
bound (\ref{opus}). For the presented parameters of $\varepsilon$ these bound values are $0.91$ and $0.36$. }
\begin{tabular}{|c||c|c|}
\hline
      & ~$\varepsilon =0.3 $~ & ~$\varepsilon = 0.8$~ \\
      \hline\hline
~${\rm dim}{\cal H}_{\rm C}=2$~ & ~0.90601~ & ~0.35990~ \\
      \hline
~${\rm dim}{\cal H}_{\rm C}=3$~ & ~0.90739~ & ~0.35994~ \\
      \hline
~${\rm dim}{\cal H}_{\rm C}=5$~ & ~0.90997~ & ~0.35996~ \\
\hline
\end{tabular}
\label{tab_1}
\end{table}

\begin{table}
\caption{The same as in Table I, but for 50 iterations. Convergence to $0.91$ and $0.36$ [these values are implied by the bound
(\ref{opus})] is seen clearly.}
\begin{tabular}{|c||c|c|}
\hline
      & ~$\varepsilon =0.3 $~ & ~$\varepsilon = 0.8$~ \\
      \hline\hline
~${\rm dim}{\cal H}_{\rm C}=2$~ & ~0.90906~ & ~0.35993~ \\
      \hline
~${\rm dim}{\cal H}_{\rm C}=3$~ & ~0.90913~ & ~0.35999~ \\
      \hline
~${\rm dim}{\cal H}_{\rm C}=5$~ & ~0.90999~ & ~0.35999~ \\
\hline
\end{tabular}
\label{tab_2}
\end{table}

\section{Summary}
\label{summa_contra_gentiles}

We studied how quantum mechanics constrains the process of transferring
density matrix elements from a system $\A$ to another system $\B$.
It was argued that the problem of matrix elements transfer lies at the
core of quantum measurements and quantum state transfer; see section \ref{intro}.

Assuming that the initial density matrix (state) $\la$ of $\A$ is
completely unknown, we show that transferring one diagonal element
$\la_{aa}$ eliminates the memory on all initial non-diagonal elements
$\la_{a\not= b}$ from the final state of $\A$.

In contrast, transferring the real part $\Re\,\lambda_{a\not =b}$ (resp.
imaginary part $\Im\,\lambda_{a\not =b}$) of a non-diagonal element
$\lambda_{a\not =b}$ eliminates the memory on $\Im\,\lambda_{a\not =b}$
(resp. $\Re\,\lambda_{a\not =b}$), and in addition the memory on the
diagonal element difference $\lambda_{aa}-\lambda_{bb}$ is eliminated.
Likewise, transferring $\lambda_{aa}-\lambda_{bb}$ eliminates the memory
on both $\Re\,\lambda_{a\not =b}$ and $\Im\,\lambda_{a\not =b}$.

Thus there is a complementarity between the diagonal and non-diagonal
elements, as well as within the triple $\Re\,\lambda_{a\not =b}$,
$\Im\,\lambda_{a\not =b}$ and $\lambda_{aa}-\lambda_{bb}$. Transferring
one element of this triple eliminates the memory on two others.
Interestingly, transferring one diagonal element implies (in general)
more severe consequences for the memory as compared to transferring
a difference between two diagonal elements.

We also studied the maximal memory that can be preserved under a
finite-accuracy [i.e., non-ideal] transfer. The proper measure of memory
is introduced in section \ref{memo} and shown to posses features
necessary for its consistent interpretation.  For each type of transfer
the maximal memory relates to the amount of non-ideality via
system-independent relations. For the transfer of non-diagonal elements
we saw that for a very inaccurate transfer, $\eps\ll 1$, the disturbance
introduced in the memory can scale as $\eps^2$, and thus can be in a
sense neglected. This is impossible when transferring diagonal matrix
elements.

Below we shall outline relations of our findings with previous results
known in literature.  Recall that transferring diagonal matrix elements
is an essential part of quantum measurement. Our relations|between the
accuracy of the diagonal elements transfer and the amount of memory
preserved for related non-diagonal elements|resemble uncertainty
relations established over the years for characterizing the information
obtained during a quantum measurement versus the induced disturbance of
the state of the measured system; see
\cite{ozawa,konrad,busch,lorenzo,martens,horod} for recent reviews on
this subject.

In the first approach (see, e.g., \cite{konrad,lorenzo,horod}) both the
information and disturbance have a global meaning. The information is
quantified, e.g., by the Shannon measure \cite{konrad,lorenzo,horod},
while for characterizing the disturbance one employs the fidelity
between the initial and final state of the measured system. The
difference with our setup is primarily that we focus on explicitly
described quantum measurements and local quantities: the quality of
measurement is determined with us by the [relative] accuracy of
transferring diagonal matrix element(s). We also use a local measure of
memory. Employing here the fidelity (or any other global measure of the
state change) will not be adequate. Moreover, as we argue in section
\ref{m_vs_f}, the fidelity does not posses some features, which are
necessary for its consistent application in this problem.

In the second approach the (des)information on the measured variable is
quantified via the overall uncertainty of the measured quantity in the
Heisenberg representation, while for characterizing the disturbance
introduced in the state of the measured system one looks at the
statistics of those variables that do not commute with the measured one;
see \cite{ozawa,busch} for reviews. This approach is well suited for
describing the Heisenberg-type uncertainty relations \cite{ozawa,busch}.
Now our approach is more flexible, because it does not
insist on doing the full measurement of the system quantity. Indeed, the
full measurement would mean transferring all diagonal elements from one
system to another. Instead, we concentrate on situations where only some
(not all) diagonal elements are transferred. Moreover, our approach
studies the transfer of non-diagonal elements that clearly goes beyond
the schemes of quantum measurements studied in \cite{ozawa,busch}. On
the other hand, we work in the Schroedinger representation and study
disturbances introduced (due to transfer) in the memory of the final
state of the source system.

With all these differences taken into account, it will be suitable to
tell that we presented a new set-up of studying information transfer
from one quantum system to another.

\section*{Acknowledgements}

It is pleasure to thank R. Balian for discussions.

The work was supported by Volkswagenstiftung.


\begin{thebibliography}{99}
\bibitem{gradient}Consider the differential $\d\tila$ of the matrix $\tila$ over an infinitesemal
increments of two independent real variables $x\equiv\Re\la_{a\not=c}$ and
$y\equiv\Im\la_{a\not=c}$ [see (\ref{togo}, \ref{kh1})]:
$$
\d\tila=\Theta_{ac}(\d x+i\d y)+\Theta^\dagger_{ac}(\d x-i\d y).
$$
To develop the notion of the matrix gradient (similar to the usual gradient)
we first calculate the Euclidean norm $||\d\tila||$ of $\d\tila$:
\BEA
||\d\tila||^2=(\d x^2+\d y^2)[\,{\rm tr}(A^2) n_x^2 ~~~~~~~~~~~~~~~~~  \nonumber\\
~~~~~~~~~~~~~+{\rm tr}(B^2) n_y^2+2\,{\rm tr}(A B) n_x n_y\,],\nonumber
\EEA
where $A\equiv\Theta_{ac}+\Theta_{ac}^\dagger$, $B\equiv i(\Theta_{ac}-\Theta_{ac}^\dagger)$, and where
$$
(n_x,n_y)=( \d x,\d y)/\sqrt{\d x^2+\d y^2}
$$
is the normal
vector over the direction of the small increment. The (modulus of) matrix gradient
is defined by maximizing over $\vec{n}\equiv (n_x,n_y)$:
$$
\frac{1}{4}{\rm max}_{\vec{n}}
||\d\tila||^2=\frac{1}{2}(\d x^2+\d y^2)[\,{\rm tr}( \Theta_{ac}\Theta_{ac}^\dagger )
+|{\rm tr}( \Theta_{ac}^2)|\,].
$$
As follows from ${\rm tr}(A B)\leq\sqrt{{\rm tr}(A^2){\rm tr}(B^2)}$, the quantity
$$
\frac{{\rm max\,}_{\vec{n}}
||\d\tila||^2}{4(\d x^2+\d y^2)}
$$ is limited from above by
${\rm tr}( \Theta_{ac}\Theta_{ac}^\dagger )=||\Theta_{ac}||^2$, and from below
by $\frac{1}{2}{\rm tr}( \Theta_{ac}\Theta_{ac}^\dagger )$. Thus, $||\Theta_{ac}||^2$
can be taken as an adequate characteristics of the matrix gradient.
\end{thebibliography}
\end{document}